# STAR Integrated Tracker

A. Rose for the STAR Collaboration,

*Wayne State University, Detroit, MI, 48203 USA*

We present the design and performance analysis of a new integrated track reconstruction code developed for the STAR experiment at RHIC. The code is meant to replace multiple previous tracker codes written in FORTRAN many years ago, and to readily enable integration of new and varied detector components. The new tracker is written from the ground up in C++ using a strong object-oriented model. Key features are an abstract geometry model for representation of detector components, a flexible track representation model, a built-in KALMAN filter for track parameter determination, and a powerful object factory model for fast handling of numerous small objects such as hits and tracks. Critical issues emphasized in the implementation of this new tracker are optimization of track reconstruction quality, minimization of reconstruction time, and memory footprint. The new tracker will be deployed and used for analysis of data acquired during the RHIC year 3 run of the STAR experiment.

## 1. INTRODUCTION

The STAR Integrated Tracker Task Force is charged with the development and implementation of a new tracking package for the STAR experiment. The interest in a new tracker spurred from the realization that the existing tracker, written in FORTRAN, was increasingly difficult to maintain, and could not readily be adapted or modified to include tracking in detectors other than the STAR TPC. It also became obvious the tracker speed would render difficult the analysis of the very large datasets the STAR experiment was about to accumulate. Moreover, the ongoing commissioning of the SVT and FTPC was bound to compound the problem, increase the complexity of the code, and its running time. A new tracker was indeed needed: one that could deliver equivalent performance in terms of track reconstruction quality, but at much increased speed, and with better maintainability and flexibility. The new code shall be written with an object-oriented design, provide for easy upgrades, addition or substitution of components.

The integrated tracker is entering the final tuning and deployment phase. The tracker is being tuned for such components as energy loss and multiple scattering, hit error parameterization, seed finder search cone size and other factors. We present here an introduction to the design an implementation of the code, as well as current reconstruction performance.

## 2. DESIGN AND IMPLEMENTATION

The new tracker is meant to provide both track finding and fitting functionality. Hits from measured with various detector components must be associated to reconstruct particle trajectories, and fitted to determine the curvature, direction, and origin of the track. One must also, and more generally, determine the momentum and species identity of the particle.

The determination of the curvature is somewhat straightforward. A minor difficulty however arises when trying to reconstruct the momentum vector of the physical particle. From a physics standpoint, the momentum vector one seeks is the vector at the vertex of origin of the particle. The problem is that the point of origin can be any of the following: a main interaction vertex, a spurious interaction vertex due to event pile-up, a secondary vertex, or a scattering center.

The track reconstruction algorithm must therefore make no a priori assumption as to the origin of the particles; the assignment of the track to a particular vertex of origin must be done after the track parameters have been determined. Viewed as an object, the track thus consists of a collection of points acquired or found with the appropriate algorithm, a parameterization of the track based on a fit of the data points to a model or template, and a vertex of origin. Properties such as the momentum (modulus or vector), and the particle identity are then calculated afterwards on the basis of the track parameters, and the known position of the vertex of origin. Note that, one can make assumptions about the vertex of origin, and include it in the fit for the determination of the track parameters after the fact, i.e. after it has been associated with the track.

One is then left with the core of the problem: finding the tracks, and fitting them to the chosen (and hopefully appropriate) track model to eventually deduce the particle final state. It then appears natural to define a "tracker" entity whose purposes are:

- To find the tracks based on a store or bank of hits reconstructed within the relevant detectors.
- To fit the hits using a suitable track model.
- To enable association with a vertex of origin and optionally allow a refit of the data including the vertex of origin.
- To calculate the final state particle information.

### 2.1. General Layout

The virtue of a Kalman Filter approach is to integrate in an efficient and compact way both the finding and fitting steps [1]. In a detector such as STAR, the track reconstruction in the Time Projection Chamber (TPC), Silicon Strip Detector (SSD), and Silicon Vertex Tracker (SVT), naturally proceeds from the outside to the inside. Track densities on outer layers of the TPC are smaller than on the inner layers, there is thus much less ambiguity in forming and following tracks. The Kalman approach enables one to progressively use the points available to refine the knowledge of the track parameters, and extrapolate (follow) the tracks inward. The calculation of the track parameters and the extrapolation from layer to layer shall proceed according to the canonical Kalman filter algorithm described here. The finder however needs a sensible seed before it can proceed in finding tracks.

Given that the number of hits in the STAR detector can be rather large for a central Au+Au collision event, it is imperative one implements a hit data store which enables fast and efficient retrieval of the relevant points. The key word is relevance. The finder shall not have to iterate on all data points to find sensible candidates for the continuation of tracks. One should thus define a measured hit/point data store, which enables point retrieval based on a layered, coarse grain pixelization of the detector.

Additionally, given that as one follows the track into the inner TPC sectors, or the SSD and SVT, ambiguity may arise as to which point is best to add on a particular track. It may thus become appropriate to fan out the tracks and follow multiple leads concurrently.

The extension of tracks from the TPC to the SVT (or backward) across structures such as the inner field cage of the TPC raises the important issue of effects caused by multiple scattering and energy losses. Given that much of the particles detected by STAR have low momenta, it is critical to include these effects properly in the propagation and fit of the tracks. We adopted much of the work done for the Alice detector by K. Safarik, and Y. Belikov [2].

The components, minimally needed, can be summarized as follows:

- Hit entities that encapsulate the position, error, energy loss, or deposition of track in detector components.
- A hit container providing polymorphic hit data storage and ultra fast retrieval of hits based on a hierarchical, layered, coarse grain representation of the detector.
- Abstract track, which define the notion of track.
- Concrete Track entities implemented following the chosen track model to hold reference to hits associated with the track, and with accessor and modifiers properties to set and get the physical properties of the track.
- A track container providing polymorphic track storage and fast retrieval based on various sorting algorithms needed, for instance, in the analysis of track merging.
- Abstract Track Finder defining the notion of tracker.
- Concrete Track Finder implementing the Kalman track finder developed in the context of this project.
- Abstract track seed finder defining the notion of track seed finder.

- Concrete Track Finder implementing a local seed finder developed in the context of this project.

## 2.2.  Tracking Algorithm

We have, in the past, explored a number of fitting algorithms for the reconstruction of tracks in a complex detector such as STAR. While global search methods based on Hough transforms, or track template may be deployed in very elegant, CPU efficient ways, and do well for the reconstruction of primary tracks, they typically do rather poorly in the reconstruction of secondary tracks – those produced from the decay of short lived particles, or from interaction within the detectors. Moreover, the application of template methods would require, for use with a detector such as STAR, a huge set of templates (even if the obvious cylindrical 12 sectors, two halves symmetry of the TPC is exploited) and would end up requiring a rather substantial memory allocation. Moreover, with such methods, as the track finding is completed, one still needs to perform a fit of the tracks that accounts for energy loss and multiple coulomb scattering effects. We have thus opted for a more conventional approach based on a Kalman filter.

We present an outline of the general track finding global strategy, track search, and fit algorithm.

### 2.2.1. Track Finding Strategy

The methodology used for the track reconstruction is basically that of a "Kalman road finder": given an existing segment of a track, use the knowledge provided by this segment, to predict and estimate where the next point on a track might be; once you got there, use the new point to update the knowledge of the track. Overall, the approach can thus be qualified as localized in space, or simply "local" by opposition to the global search techniques alluded to in the introduction of this section.

STAR uses the notions of global, primary, and secondary tracks. Primary tracks are those emanating directly from the main collision vertex whereas secondary tracks are produced by decay or interaction of primary tracks within the detector. The finite resolution of the track reconstruction, and kinematical focusing of decay products concur to render the distinction between many secondary and primary tracks rather difficult. STAR thus first analyze all tracks as if they were secondary tracks, and do not include the main collision vertex. One then search for the fraction of those that present a good match with the main collision vertex and can be labeled as primaries. The tracks obtained in the first pass are labeled "global tracks" and are fitted without a vertex. The primary tracks are extension of the global tracks including the vertex: their fit includes the vertex. Note that STAR maintains a double list of tracks consisting of global and primary tracks, where tracks that match the main vertex appear twice - once as global and once as primary. It is thus possible to recover the track parameters with and without the primary vertex for further analysis of V0s and other decay topologies.

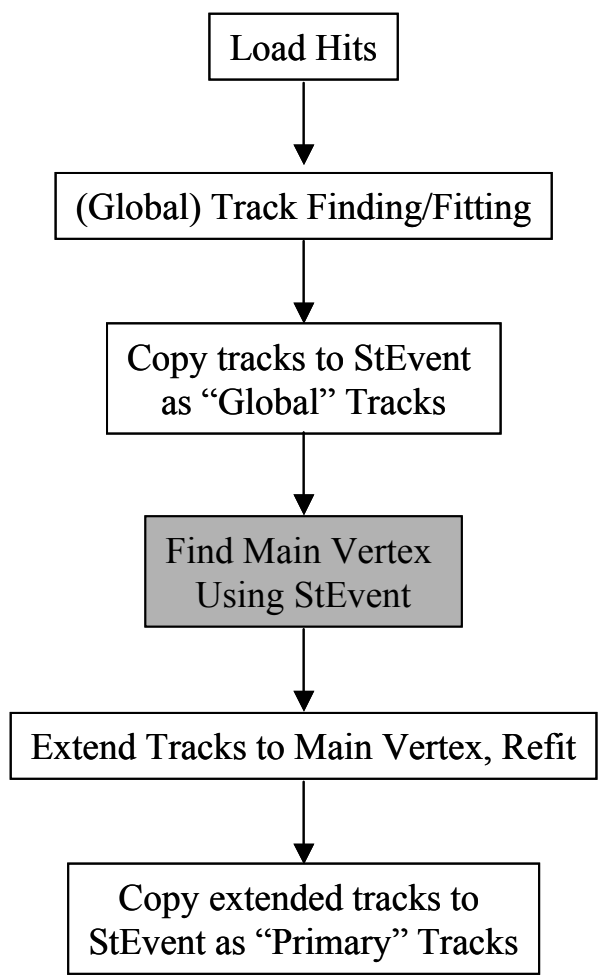

Figure 1: General Track Reconstruction Strategy. Sequence of tasks involved in the track reconstruction. Note that the main vertex is outside the scope of this project.

The persistent data model for STAR (StEvent) is a class containing a single event and its characteristics. This event model also contains a track model called StTrack. As we started to develop this new tracker, we felt the STAR StTrack model did not provide the flexibility and efficiency need for the tracker, and we thus designed and implemented a new track model for use within the new framework. Given that much of the existing STAR C++ code already use the StTrack model, we concluded it would be simpler to keep the existing track model for i/o purposes while conducting the track search with the StiTrack model. This implies that once StiTrack tracks have been found, they must be copied into the StEvent format.

The track search and event reconstruction algorithm, proceeds in five basic steps. The first step consists in the actual track search and is described in the following section. It produces so called "global tracks", or tracks with no association to the primary vertex. Those global tracks are then copied into the STAR event model StEvent/StTrack by a call to a filler helper class method. The main vertex finder is called next (with StEvent as argument) to find the vertex of the event. If a vertex is found, the Kalman vertex finder is called, once again, to attempt an extension of all found tracks to the main vertex. The event filler is then called once more to copy the newly found primary tracks, i.e. those tracks that were successfully extended to the main vertex. The track reconstruction is then completed.

### 2.2.2. Track Search and Fitting Algorithm

Tracking proceeds in two steps: candidate, or "seed", finding and track extension and fitting.The search first uses a Kalman road finder to collect track candidates and proceeds to extend these candidates sequentially until no more tracks are found. No correlations between tracks are considered although hits may initially belong to more than one track.

The search for each track is initiated with a call to a Track Seed Finder. The search stops when the seed finder returns no seed. Track seeds are short track stubs consisting of a sequence of a few hits. As such, they carry just enough information to enable a very rough estimate of the track position, direction, and curvature. Seeds returned by the seed finder are not confined to any specific region of the detector. However, in the case of the STAR detector it is easier to find reliable track patterns in a low track density environment, so the search for seeds proceeds from the outside in. Therefore, the seeds returned are typically located near the periphery of the detector.

The rough estimate of the track provided by the seed is used by the Kalman finder to begin the extension and search of the track through the detector. Since the seeds predominantly lie near the periphery of the detector, the Kalman search that follows first proceeds inward. The Kalman-search proceeds through the virtual layers of the detector, step by step. It is considered complete when the search reaches the inner most volume, or when a prescribed minimum number of active detector layers have been crossed without finding matching hits. The mathematical details of the Kalman search and fit are described in the detailed documentation of this project on the STAR web site. The Kalman finder uses the direction and curvature of the existing track stub to estimate (extrapolate) the position of the next track hit on the next available layer.

Matching hits are then sought on that layer within a radius of confidence determined by the error parameters of the track. If no matching hit is found, the given layer is skipped. If one or more matching hit candidates, one calculates the increment of track chi-square caused by the addition of the candidate hits. Candidates are deemed acceptable if the chi-square increment is smaller than a prescribed (user settable) maximum. If more than one candidate hit satisfies the chi2 requirement, one selects and adds to the track the hit with the lowest incremental chi-square value. Once a hit is added, the track parameters (i.e. curvature, direction, etc) are updated using the Kalman track model. As the track-search proceeds inward and eventually reaches the inner most detector volume, the track parameters are progressively refined. The Kalman parameters (including the chi-square) of the track at the last hit are the best estimator of the track.

Given that the track search initially proceeds on the basis of a seed that may lie deep inside the detector, it is possible that the inward finding and fitting pass might result in an incomplete track. Examination of the outermost point of the track determines if the track should be extended outward toward the edge of the detector. The search is considered complete if a number of points smaller than a prescribed minimum could be added, and the tracked proceeds to the next seed. If the track can be extended, the continuation of the track outward proceeds similarly to the inward pass. Successive virtual layers are search step by step for additional hits, and the track parameters are updated at each step. Note however that in order to initiate the outward pass, an outward refit of the track is first performed in order to update the track parameters of the outer most node of the track. The fit is performed with the same machinery (methods) than those used by the finder. The only difference lies in the fact that the hits are already found, so one only needs to update the track parameters. The outward search proceeds until the edge of the detector or until too many layers have been crossed without association of hits on to the track. The same threshold is used here as for the inward pass.

If an outward pass is performed, and once completed, the track parameters of the inner track nodes can be considered under constrained since not all hits on the track were used to calculate the track parameters for those nodes. An inward track refit is thus accomplished.

If an outward pass is not performed, the track parameters of the outer nodes can also be considered under constrained. An outward final fit is thus conducted. This fit is deemed necessary to provide best track parameter knowledge on the outset of the track, which may then be used by user analyses for extension of the tracks to non-tracking detectors such as, in STAR, the CTB, the TOF, or the EMC.

## 2.3. Deployment and Running Conditions

Reconstruction of STAR data is typically done at the RHIC Computing Facility (RCF). Each node in the computer bank contains two CPU's sharing 1 gigabyte of memory. Previously, reconstruction has been hindered by process memory leaks, which swell the size of the executable binary greater than 500 megabytes; thus preventing more than one job per node and reducing the efficiency of the computer farm. This situation required supervision and occasional intervention by the analysis team to maintain efficiency. Typically, individual reconstruction had to be limited to small set of events (<100), to curtail memory overrun.

The new tracking framework, through extensive use of standard packages (such as the Standard Template Library) and coding practices, has achieved little or no memory leak at runtime, with a maximum memory footprint significantly below the level of the previous tracker. A snapshot of the memory footprint for the tracking code was taken as 300 central events were reconstructed. The memory size (in Mega-bytes) is displayed in Figure 1 as a function of the event number analyzed. A sharp rise can be seen as the tracker allocates new blocks of memory to accommodate larger events, but a plateau is reached as it reuses these memory blocks for subsequent events.

We stress that we were able to operate the code to analyze multiple thousands events without crash, segmentation faults, or increase of the memory footprint. This fact in itself will be a tremendous improvement over the previous incarnation of the STAR tracking code. We also stress that the code's memory footprint of less than 400 Mbytes enables efficient concurrent use of the dual CPU computers with 1 GB of available memory, located at BNL, LBNL, Wayne State University group, and other STAR institutions.

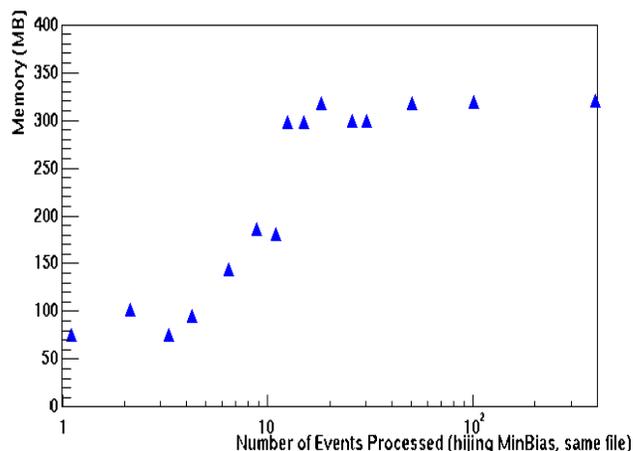

Figure 2: Memory footprint vs. Event number during reconstruction. The increase in memory consumption as larger events are encountered can be seen.

We also have paid careful attention to the speed performance of the code. The code was designed to avoid repeated unnecessary operations such object instantiation and deletion. Data structures are designed to allow fast retrieval of the relevant information. We also avoided the use of string based searches, etc. The code is thus rather fast. The time performance is displayed in Figure 1b which shows a plot of the total tracking time as a function of the number of tracks in each event.

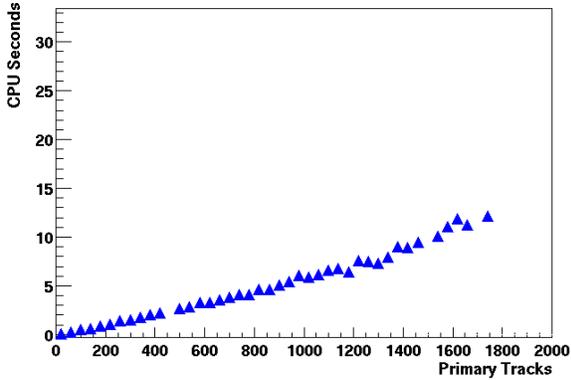

Figure 3: Total tracking time as a function of the number of tracks reconstructed. CPU seconds is in arbitrary units, which will vary depending on the speed of the computer used.

The tracking time scales linearly with the number of track reconstructed. Central collision events are reconstructed in less than 15 CPU seconds on a typical RCF node. This corresponds to an improvement of a factor of 6 relative to the previous tracker which should be extremely beneficial for the analysis of future production of STAR data given that it is foreseen the data volume accumulated each will substantially increase thanks to improvement to the STAR data acquisition system. It shall then be possible to analyze the data faster and possibly through multiple iterations as needed.

## 3. PERFORMANCE CHARACTERIZATION

We present below a brief summary of the performance of the tracker. The basic performance characteristics detailed here are hit finding efficiency, track reconstruction efficiency and momentum resolution.

### 3.1. Hit Association Efficiency

An essential measure of the performance of the tracker is the efficiency of associating related hits into a track. Figure 2 presents a study of the hit association efficiency of the tracker based on simulated (Monte Carlo) events. The plot shows in ordinate the average ratio of the number of found hits (i.e. associated to a track) to the number of hits belonging to a MC track as a function of the total charged particle multiplicity (in an arbitrary but fixed angular acceptance) of the events. The ratio of found hits to MC hits peaks at ~80% for low multiplicity events (peripheral collisions) and decreases monotonically for increasing event multiplicity. The rather modest value of 80% arise in part because of losses at sector boundaries, and in part due to hit losses in low pt track with segments nearly parallel to the TPC pad planes. The monotonic decrease occurs due to the increased space point occupancy in more central, higher track multiplicity events.

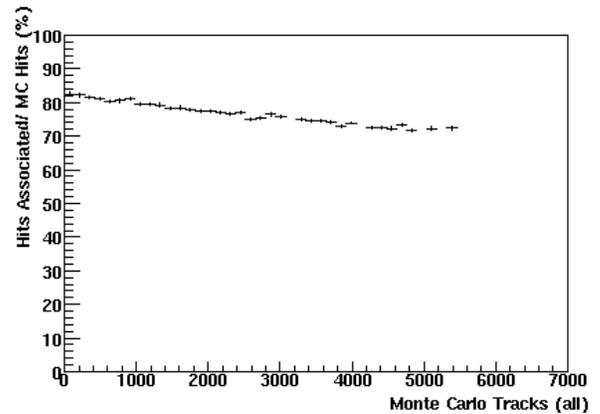

Figure 4: Average Hit Association efficiency as a function of the event multiplicity. The hit association efficiency is defined as the ratio hits properly associated to the number of hits on the track.

### 3.2. Track Reconstruction Performance

We next consider the overall track reconstruction efficiency as a function of the track transverse momentum. Figure 5 presents a study based on HIJING simulated events of the charged pion track reconstruction efficiency as a function of the track transverse momentum. The efficiency is defined as the number of tracks reconstructed with more than 15 hits (for MC tracks which also have more than 15 hits). The efficiency achieved with the ITTF tracker is consistently high (85-90%) for transverse momenta above 0.5 GeV/c. Softer tracks, however, are reconstructed with a lower efficiency by the ITTF tracker. These low momentum tracks are typically lost due to scattering in the material of the detector.

### 3.3. Transverse Momentum Resolution

The resolution of the reconstructed transverse momentum, shown in Fig. 6, reaches a minimum at 500 MeV/c. This minimum (1.2%), although already quite good, is expected to improve with further tuning.

The resolutions are very sensitive to the corrections for energy loss of the track in the detector media. These corrections involve detailed knowledge of the radiation length the track transverses, and so are dependant on an accurate estimate of the type, thickness, and placement of materials in the detector.

Inaccurate understanding of the materials will result in a bias to the measured momentum. Currently, the bias for the integrated tracker is small, but still significant for the lower momentum tracks (2% bias at 200 MeV/c). Work is proceeding to better reconstruct the materials traversed. Currently, this correction is accomplished through parameters describing the composition and position of the detectors set in the tracking code itself. Future plans include extracting this information from an existing online database used in simulation.

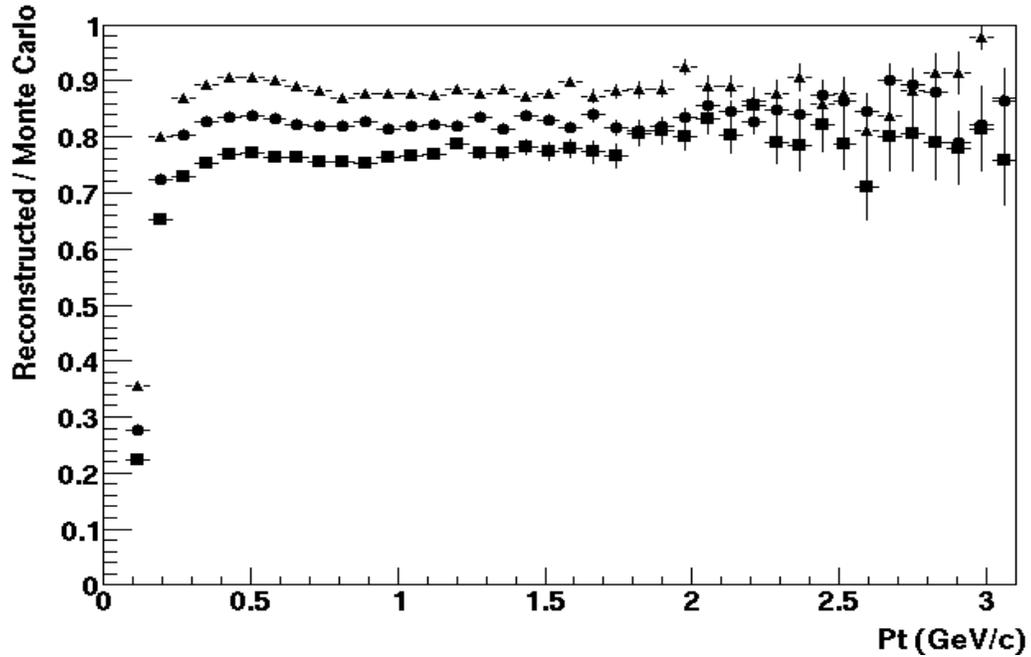

Figure 5: Track reconstruction efficiencies of the tracker as a function of the track transverse momentum. Efficiency is calculated as the ratio between the number of reconstructed tracks matched to an input Monte Carlo track and the number of Monte Carlo tracks within the detector acceptance. Triangles represent data from the most peripheral collisions, circles from intermediate centralities, and squares from the most central.

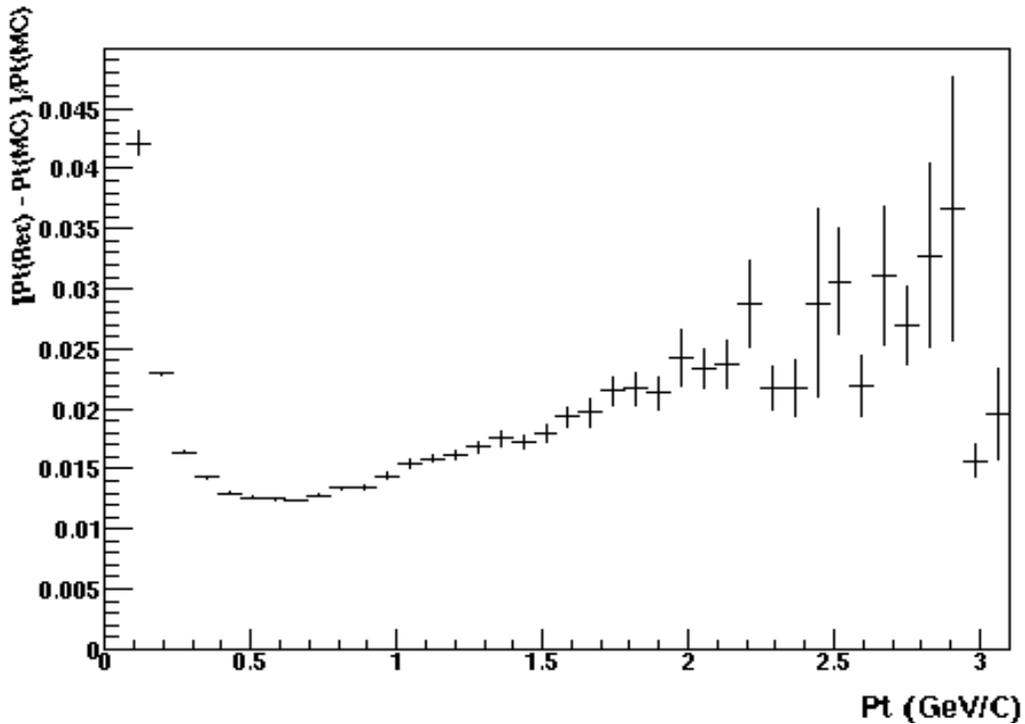

Figure 6: Transverse Momentum Resolution as a function of transverse momentum for all tracked particles. The best momentum reconstruction is achieved for tracks with a transverse momentum of .6 GeV/c, with a resolution of 1.2.

## 4. CONCLUSIONS

The STAR Integrated tracker has shown very positive initial results and performance. The implementation of the Kalman track finding and fitting algorithm has been validated. The code shows significant speed and stability improvements over the previous FORTRAN-based software package. To match track kinematic reconstruction performance goals, set by the current software, tuning and optimization of tracking parameters must be studied.

## Acknowledgments

This work is supported through the Department of Energy, contract DE-FG02-92ER40713.